\documentclass{aastex631}

\newcommand{\um}{$\mu$m}
\newcommand{\kms}{km~s$^{-1}$}
\newcommand{\hinode}{\textit{Hinode}}
\renewcommand{\ion}[2]{#1\,\textsc{#2}}
\newcommand{\as}{$^{\prime\prime}$}
\newcommand{\lam}{$\lambda$}

\begin{document}

\title{Updated reference wavelengths for  Si\,VII and Mg\,VII lines in the 272--281 Angstrom range}

\author[0000-0001-9034-2925]{Peter R. Young}
\affiliation{NASA Goddard Space Flight Center, 
Greenbelt, MD 20771, USA}
\affiliation{Northumbria University, Newcastle upon Tyne, NE1 8ST, UK}



\begin{abstract}
New reference wavelengths for atomic transitions of \ion{Mg}{vii} and \ion{Si}{vii} in the 272--281~\AA\ wavelength range are derived using measurements from the Extreme ultraviolet Imaging Spectrometer (EIS) on board the \hinode\ spacecraft. 
\ion{Mg}{vii} and \ion{Si}{vii} are important ions for measuring plasma properties in the solar transition region at around 0.6~MK. The six \ion{Si}{vii} wavelengths are 13--21~m\AA\ and 7--11~m\AA\ longer than the values in the NIST Atomic Spectra Database (ASD) and the compilations of B.~Edl\'en, respectively. The four \ion{Mg}{vii} wavelengths are shorter than the values in the ASD by 8--12~m\AA\ but show reasonable agreement with the Edl\'en values.
The new wavelengths will lead to more accurate Doppler shift measurements from the EIS instrument, and will be valuable for spectral disambiguation modeling for the upcoming Multi-Slit Solar Explorer mission.
\end{abstract}

\section{Introduction} \label{sec:intro}

\ion{Mg}{vii} and \ion{Si}{vii} are formed at a temperature of 0.6~MK in electron-ionized plasmas and give rise to 10 emission lines in the 272--281~\AA\ wavelength range. These lines have been observed by the Extreme ultraviolet Imaging Spectrometer \citep[EIS:][]{2007SoPh..243...19C} on board the \hinode\ spacecraft since 2006, and they provide valuable diagnostics of the upper transition region. For example, the \ion{Mg}{vii} \lam280.73/\lam278.39 ratio is an excellent diagnostic of the electron number density \citep{2007PASJ...59S.727Y} that has been used to obtain the density variation along coronal loops \citep{2009ApJ...694.1256T,2012ApJ...744...14Y}. The \ion{Si}{vii} 275.37~\AA\ line is often used for Doppler velocity measurements \citep{2011ApJ...730...37U} and is an important ``anchor line" for accounting for the time-variable wavelength offset between the short and long wavelength channels of EIS \citep{2012ApJ...744...14Y}.

\citet{2009ApJ...706....1L} highlighted that there are significant discrepancies  for the \ion{Mg}{vii} and \ion{Si}{vii} reference wavelengths between the tabulations in the National Institute for Standards and Technology (NIST) \href{https://www.nist.gov/pml/atomic-spectra-database}{Atomic Spectra Database} (ASD) and those provided by B.~Edl\'en \citep{1983PhyS...28...51E,1985PhyS...31..345E}. The present work provides updated reference wavelengths for these lines based on measurements obtained from the EIS spectra.

All data and software used in this article are publicly available, and Section~\ref{sect:avail} gives the sources. Section~\ref{sect:lines} presents the \ion{Mg}{vii} and \ion{Si}{vii} lines studied here, and Section~\ref{sect:cool} gives some properties of the cool coronal loops that are analyzed. Section~\ref{sect:eis} describes the EIS instrument and explains the choice of datasets, and Section~\ref{sect:analysis} details the data analysis procedure used to derive the new reference wavelengths. Section~\ref{sect:lopt} gives new experimental energies obtained from the wavelengths, and a summary of the article is given in Section~\ref{sect:summary}.

\section{Data and software availability}\label{sect:avail}

Data from the \hinode/EIS instrument are analyzed in this article and the level-0 (uncalibrated) data are publicly available from the \href{https://sdac.virtualsolar.org/cgi/search}{Virtual Solar Observatory} and the \href{http://sdc.uio.no/sdc/}{Hinode Science Data Centre Europe}. The EIS data are calibrated using IDL software available in the public repository \emph{Solarsoft} \citep{1998SoPh..182..497F,2012ascl.soft08013F}, which also contains routines for extracting spectra and fitting Gaussians as described in the following text. Derived data products from the present analysis are made available through Zenodo \citep{young_peter_r_2023_8368508} and include Gaussian line fit parameters from which the emission line wavelengths are derived. 
The data files used as input to the LOPT code (Section~\ref{sect:lopt}) are available through a separate Zenodo entry \citep{young_2023_10001618}.
IDL routines for generating the figures in this article are available through a \href{https://github.com/pryoung/papers}{GitHub repository} in the folder \textsf{2023\_mg7\_si7}.

\section{The Mg VII and Si VII lines}\label{sect:lines}

\ion{Mg}{vii} belongs to the carbon isoelectronic sequence and the EIS wavelength ranges contain four emission lines corresponding to the $2s^22p^2$ $^3P_J$ -- $2s2p^3$ $^3S_1$ ($J=0,1,2$) and $2s^22p^2$ $^1D_2$ -- $2s2p^3$ $^1P_1$ transitions with wavelengths between 276~\AA\ and 281~\AA. Of the $^3P_J$--$^3S_1$ transitions, only the weakest line at 276.14~\AA\ is unblended. The strongest line at 278.39~\AA\ is blended with \ion{Si}{vii}  278.46~\AA, but can be resolved through a two-Gaussian fit, for example. The $^3P_1$--$^3S_1$ transition at 276.99~\AA\ is blended with two \ion{Si}{viii} lines and it is not possible to resolve this feature without using information from one of the other \ion{Mg}{vii} lines \citep{2009ApJ...706....1L}. The $^1D_2$--$^1P_1$ transition at 280.73~\AA\ is unblended and makes an excellent density diagnostic with either of the 276.14~\AA\ or 278.39~\AA\ lines \citep{2007PASJ...59S.727Y}.

\ion{Si}{vii} is a member of the oxygen isoelectronic sequence and the six members of the $2s^22p^4$ $^3P_J$ -- $2s2p^5$ $^3P_{J^\prime}$ ($J=0,1,2$; $J^\prime=0,1,2$) multiplet are found between 272 and 279~\AA, close to the \ion{Mg}{vii} lines discussed above. The 0--1 transition occurs at 276.86~\AA\ and is blended with two \ion{Si}{viii} lines. As for the \ion{Mg}{vii} line mentioned above, it is not possible to resolve this blend without using information from one of the other \ion{Si}{vii} lines, and so it is not useful for the present analysis. The 1--2 transition is partially blended with a stronger \ion{Mg}{vii} line, as discussed above. The 1--0 transition at 274.19~\AA\ is blended with \ion{Fe}{xiv} 274.20~\AA, but is important as the only line that  determines the $2s2p^5$ $^3P_{0}$ energy. The multiplet's lines show density sensitivity relative to each other and they are most sensitive around $2\times 10^8$~cm$^{-3}$, but this is too low for structures in the low solar atmosphere.

\section{Cool loops}\label{sect:cool}

\begin{figure}[t]
    \centering
    \includegraphics[width=0.8\textwidth]{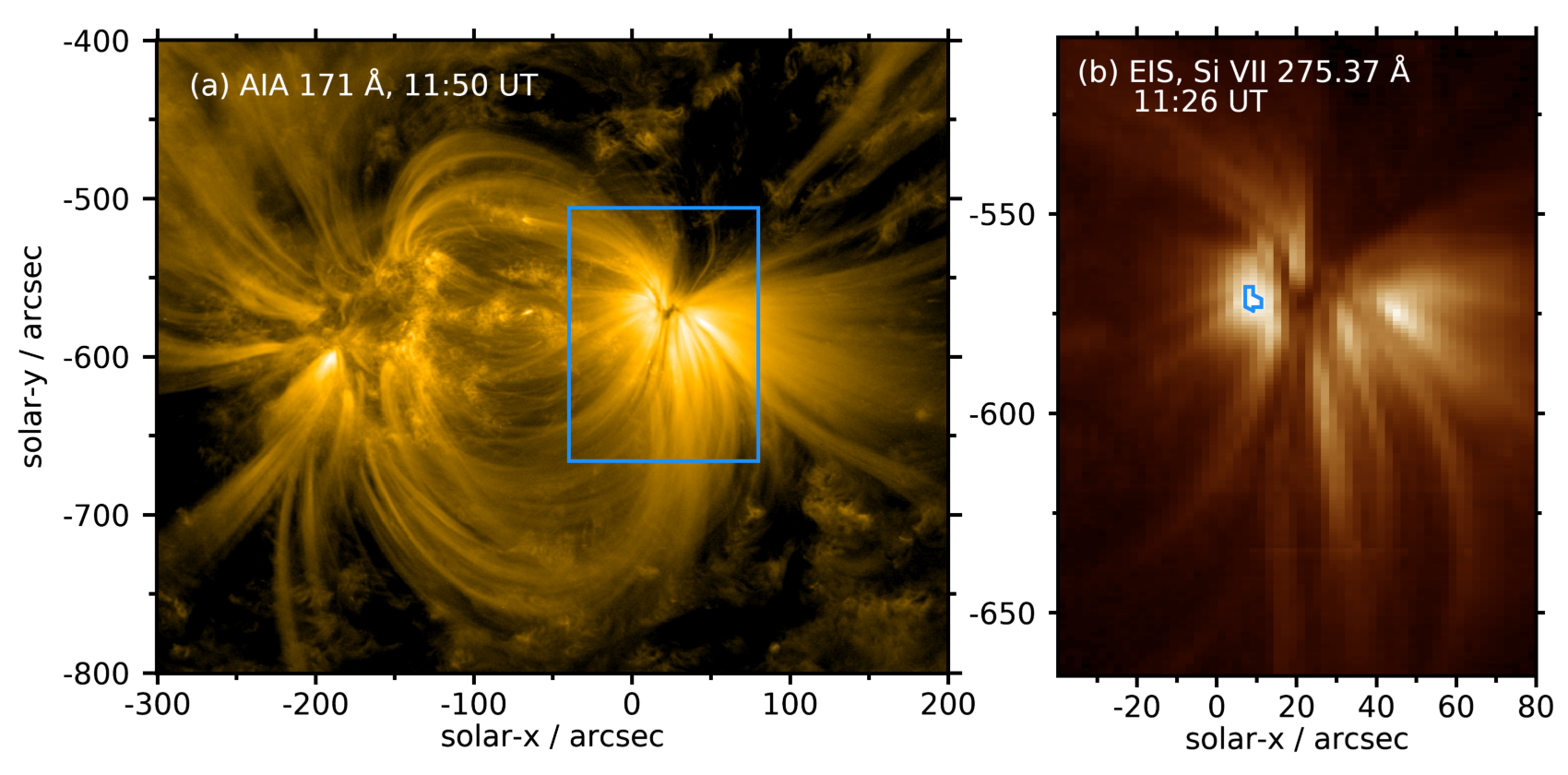}
    \caption{(a) an AIA 171~\AA\ image (logarithmic scaling) showing an active region from 2010 September 22. The blue box shows the region rastered by EIS. (b) an EIS  image (linear scaling) formed from the \ion{Si}{vii} 275.37~\AA\ line observed in the raster beginning 11:26~UT. The blue contour shows the spatial region averaged to yield the spectrum from which wavelength measurements were made in the present work. Three adjacent data columns between $x=66$\as\ and $x=72$\as\ are missing in the EIS data and have been interpolated for display purposes.}
    \label{fig.image}
\end{figure}

Solar active regions often have  loops at their peripheries that are rooted in decaying plage or sunspots. These loops are bright in \ion{Fe}{ix} and \ion{Fe}{x}, formed around  0.8--1.0~MK, and they came to prominence when the Transition Region and Coronal Explorer \citep{1999SoPh..187..229H} began taking high resolution images in a filter centered at 173~\AA. These images were dominated by emission from \ion{Fe}{ix} 171.07~\AA\ and \ion{Fe}{x} 174.53~\AA. The loops are also observed in images from the Atmospheric Imaging Assembly \citep[AIA:][]{2012SoPh..275...17L} on board the Solar Dynamics Observatory which has a channel centered at 171~\AA\ that is dominated by \ion{Fe}{ix}. Figure~\ref{fig.image}a shows an AIA 171~\AA\ image for an active region studied in the present work, observed on 2010 September 22. As the loops are cooler than the shorter loops in the active region center, they are often referred to as ``cool"  loops.

For ions formed at lower temperatures, the emission is increasingly concentrated at the footpoints of the cool loops (Figure~\ref{fig.image}b) and has been studied with the Coronal Diagnostic Spectrometer \citep{1995SoPh..162..233H} on board the Solar and Heliospheric Observatory, and \hinode/EIS. In particular, the emission lines of ions of magnesium, silicon and iron are strongly enhanced in the temperature range 0.4--0.8~MK: see Figure~4 of \citet{2007PASJ...59S.727Y}. Spectral atlases and detailed plasma diagnostics of cool loop footpoints have been presented by \citet{2009ApJ...706....1L}, \citet{2009ApJ...707..173Y} and \citet{2009A&A...508..501D}. The high signal-to-noise of cool loop footpoint spectra make them ideal for obtaining reference wavelengths for lines in the 0.4--0.8~MK temperature range, and they have been used recently for re-assessments of energy levels of \ion{Fe}{vii} \citep{2021ApJ...908..104Y,2022ApJS..258...37K} and \ion{Fe}{ix} \citep{2022ApJ...936...60R}.

\section{The EIS instruments and datasets}\label{sect:eis}

EIS is an imaging slit spectrometer that obtains spectra in the 170--212~\AA\ and 246--292~\AA\ wavelength ranges at a resolution of 3000--4000. Only data from the long wavelength (LW) channel are used in the present work. There is a choice of four slits with different widths, and only data from the narrow 1\as\ and 2\as\ slits are used here. The instrument spatial resolution is 3--4\as\ \citep{2022SoPh..297...87Y}, and the detector pixels correspond to an angular size of 1\as.

Thirteen EIS datasets were used for the present analysis. All are spatial rasters that yield the entire spectral range of the EIS instrument, and all are active region datasets in which cool loop footpoints are prominent. The file IDs for the datasets are given in Table~\ref{tbl.files}. The first eight characters give the date in year-month-day format, and the last four characters give the start time of the observation in hours and minutes, e.g., ``2152" corresponds to 21:52~UT. The dataset 20070105\_2152 was previously studied by \citet{2009A&A...508..501D}, and the dataset 20070221\_0112 was previously studied by \citet{2009ApJ...706....1L} and \citet{2009ApJ...707..173Y}, although the spectra derived here were obtained from different spatial regions. 

The datasets were selected either from those already known by the author, or by browsing the \href{https://eismapper.pyoung.org}{\textit{EIS mapper}} \citep{young_peter_r_2022_6574455} products that show the locations of EIS rasters on full-disk solar EUV images. For the latter, rasters were sought that included bright loop footpoints as seen in the AIA 171~\AA\ channel. Inspection of selected datasets was performed to avoid footpoints for which there was significant hot (1--3~MK) emission in the line-of-sight in order to minimize blending. There was a particular focus on datasets from 2007 when EIS was  in a near-pristine condition. In particular, the number of warm pixels on the EIS detectors was low and the sensitivity of the long-wavelength channel was high.

The \ion{Si}{vii} 274.19~\AA\ line is blended with \ion{Fe}{xiv} 274.20~\AA\ (formed at 2~MK). Potentially this can be resolved by considering data from an on-disk coronal hole, where \ion{Fe}{xiv} is expected to be negligible due to the low temperature in coronal holes \citep[e.g.][]{1999JGR...104.9753D}. However, instrument scattered light \citep{2018ApJ...856...28W,2022ApJ...938...27Y} results in a baseline intensity component to hot emission lines within coronal holes and it was determined that the cool loop footpoints are preferable as locations where \ion{Si}{vii} dominates \ion{Fe}{xiv} at 274.2~\AA.

The spectra from each dataset were prepared as follows. An image from the raster was formed in the \ion{Si}{vii} 275.37~\AA\ line, and a location was chosen where the line emission was strong. A number of spatial pixels around this location were averaged to yield a single spectrum (e.g., Figure~\ref{fig.image}b). The number of pixels selected varied from 10 to 32. The averaging was done using the IDL routine \textsf{eis\_mask\_spectrum} \citep{young_peter_r_2022_6339584}, and the output spectra are available at Zenodo \citep{young_peter_r_2023_8368508}.

\section{Data analysis}\label{sect:analysis}

Deriving absolute reference wavelengths for EUV emission lines using solar spectra requires some assumptions as plasma motions can lead to Doppler shifts of tens of m\AA. In addition, EUV instruments cannot be flown with wavelength calibration lamps and so the absolute wavelength scale is not known. A common practice is to assume that  measurements made above the solar limb in a quiet part of the solar atmosphere represent plasma with zero Doppler shift. This is because the solar magnetic field is approximately parallel to the Sun's radius and plasma is constrained to move along the magnetic field direction in the low corona. Thus, motions will be mostly perpendicular to the line-of-sight at the limb and will not give rise to Doppler shifts. In addition, the long plasma column depth above the limb serves to smooth out any small spatial scale plasma motions.

The quiet Sun off-limb spectrum thus gives a zero Doppler shift spectrum, but there remains the problem of the unknown absolute wavelength calibration. For this, the \citet{1976ApJ...203..521B} rocket spectrum is used. This high resolution, full-disk solar spectrum was calibrated using known wavelengths of various cool emission lines in the solar spectrum. By adjusting lines in the EIS spectrum to be consistent with those in the \citet{1976ApJ...203..521B} spectrum the absolute wavelength scale can be set.

This method potentially allows all of the \ion{Si}{vii} and \ion{Mg}{vii} lines in the EIS spectrum to be measured. However, these lines are weak in off-limb regions which typically have temperatures around 1.0--1.5~MK \citep{2009ApJ...700..762W}, and only the \ion{Si}{vii} 275.37~\AA\ line can be reliably measured. The EIS off-limb measurements of \citet{2011ApJ...727...58W} are used, which include the \ion{Si}{vii} line as well as the \ion{Fe}{xiv} 274.20~\AA\ and \ion{Fe}{xv} 284.16~\AA\ lines. The latter are in the \citet{1976ApJ...203..521B} line list with an accuracy of 4~m\AA. (The \ion{Si}{vii} line is also present but with a low accuracy of 10~m\AA.) An offset to the \citet{2011ApJ...727...58W} wavelengths is therefore obtained by adjusting the \ion{Fe}{xiv}  and \ion{Fe}{xv} lines to best match the \citet{1976ApJ...203..521B} wavelengths. This offset then gives a \ion{Si}{vii} reference wavelength of 275.368~\AA\ with an uncertainty of 4~m\AA, which is the uncertainty of the \citet{1976ApJ...203..521B}  coronal line measurements.

With the \ion{Si}{vii} 275.37~\AA\ reference wavelength obtained, the high signal-to-noise cool loop footpoint spectra are then used to derive wavelength offsets for the remaining \ion{Si}{vii} and \ion{Mg}{vii} lines relative to this line. Since CHIANTI predicts these two ions to be formed at the same temperature, they are expected to display the same Doppler shifts and so there should be no relative offset between the two ions. This is confirmed here by both Doppler shift and intensity measurements (see below).

\begin{figure}[t]
    \centering
    \includegraphics[width=\textwidth]{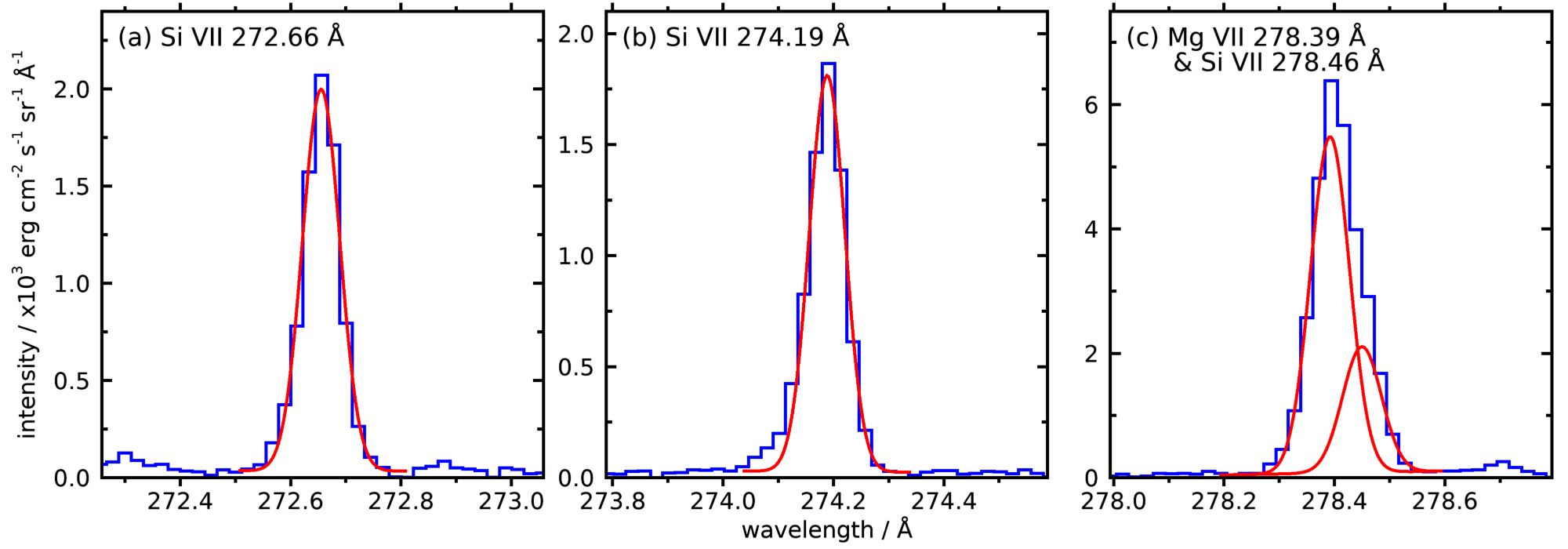}
    \caption{Line profiles from the 20070221\_0115 dataset. The spectra are shown in blue and Gaussian fits to the lines in red. Panel (c) shows the two-Gaussian fit for this blended feature.}
    \label{fig.fits}
\end{figure}

The wavelength offsets were obtained as follows. For each of the 13 cool loop spectra, emission lines were manually fit with Gaussians using the \textsf{spec\_gauss\_eis} routine \citep{young_peter_r_2022_6339584}.
Of the 10 \ion{Mg}{vii} and \ion{Si}{vii} lines, two are not analyzed here as they are blended (Section~\ref{sect:lines}). Four of the lines are isolated and could be fit with single Gaussians. The \ion{Si}{vii} 272.66~\AA\ line has an unknown, weak line on the short-wavelength side that was masked out from the fit (Figure~\ref{fig.fits}a). The blend at 274.2~\AA\ containing a \ion{Si}{vii} line and an \ion{Fe}{xiv} line was fit as a single Gaussian, however there is an unknown, weak line on the short-wavelength side apparent in some spectra (Figure~\ref{fig.fits}b) that was fit with a second Gaussian. As noted earlier, the \ion{Mg}{vii} 278.39~\AA\ and \ion{Si}{vii} 278.46~\AA\ lines are partially blended and were fit together using two Gaussians forced to have the same width (Figure~\ref{fig.fits}c). The latter is justified as the two ions are emitted at the same temperature, hence their thermal widths will be very similar. 

In addition to the above lines, a two-Gaussian fit was applied to the blend of \ion{Mg}{vi} 270.39~\AA\ and \ion{Fe}{xiv} 270.52~\AA. The intensity of the latter is used to estimate the strength of \ion{Fe}{xiv} 274.20~\AA, which blends with one of the \ion{Si}{vii} lines. The \lam274.20/\lam270.52 ratio is weakly density sensitive with a value $\le 2$ for densities above $10^9$~cm$^{-3}$, which are typical of the active region corona. Therefore, the 270.52~\AA\ intensity is multiplied by two to give an upper limit intensity for the \ion{Fe}{xiv} 274.20~\AA\ line. 

For each of the \ion{Si}{vii} and \ion{Mg}{vii} lines except \ion{Si}{vii} 274.19~\AA, the measured wavelengths were used to derive the wavelength separations relative to \ion{Si}{vii} 275.37~\AA. The average separation, $\Delta\lambda$, for each line is shown in Table~\ref{tbl.sep}. The uncertainty is obtained from the standard deviation of the 13 measured values.

For \ion{Si}{vii} 274.19~\AA\ there is a risk that the the \ion{Fe}{xiv} blend skews the wavelength measurement. To minimize this, the intensity ratio of the measured line relative to \ion{Si}{vii} 275.37~\AA\ is computed and only datasets with ratios $\le 0.4$ are retained. The theoretical \ion{Si}{vii} \lam274.19/\lam275.37 ratio from CHIANTI \citep{2023arXiv230515221D} ranges from 0.21 to 0.26 over the density range $10^9$~cm$^{-3}$ to $10^{10}$~cm$^{-3}$ expected for the cool loop footpoints, hence the requirement is to ensure a dominant \ion{Si}{vii} contribution. In addition, the \ion{Fe}{xiv} 270.52~\AA\ intensity is multiplied by two, and then divided by the measured 274.19~\AA\ intensity. Only datasets for which this ratio is $< 0.5$ are retained. Seven datasets satisfy these two criteria, and the wavelength differences are averaged to yield the $\Delta\lambda$ value in Table~\ref{tbl.sep}.

\begin{deluxetable}{ll}[t]
\tablecaption{Identifiers for the EIS data files used in the present analysis.\label{tbl.files}}
\tablehead{
   \multicolumn{2}{c}{File ID} }
\startdata
20070105\_2152 & 20070106\_1555 \\
20070115\_2217 & 20070123\_1540 \\
20070221\_0015 & 20070221\_0112 \\
20070524\_1229 & 20100801\_2339 \\
20100809\_0341 & 20100922\_1126 \\
20110122\_1512 & 20160520\_0050 \\
20170713\_1251 &  \\
\enddata
\end{deluxetable}

The $\Delta\lambda$ standard deviations for the two unblended \ion{Mg}{vii} lines are small, translating to $< 1.5$~\kms\ in Doppler motions. This confirms the assumption that \ion{Mg}{vii} is formed very close in temperature to \ion{Si}{vii} and thus they exhibit very similar Doppler motions in solar plasmas. This is backed up by the intensity ratio of \ion{Mg}{vii} 276.14~\AA\ to \ion{Si}{vii} 275.37~\AA, which shows little variation between the datasets. It lies between 0.17 and 0.21 (mean value of 0.186) for 12 of the datasets, with the 20070123 dataset  anomalous in having a ratio of 0.26. Although not directly relevant to the present article, it is to be noted that the CHIANTI atomic models combined with the \citet{2021A&A...653A.141A} solar photospheric abundances predict a \lam276.14/\lam275.37 ratio of 0.119. The higher ratios found in the cool loop footpoints may be due to an enhanced magnesium abundance (by around 50\%).

\begin{deluxetable}{lllccccc}[t]
\tablecaption{\ion{Si}{vii} and \ion{Mg}{vii} transitions and wavelengths. The Ritz wavelengths are the recommended values.\label{tbl.sep}}
\tablecolumns{8}
\tablehead{
  &&&&\multicolumn{4}{c}{Wavelengths/\AA}\\
  \cline{5-8}
  \colhead{Ion} &
  \colhead{Configurations} &
   \colhead{Terms} &
   \colhead{$\Delta\lambda$\tablenotemark{a}/\AA} &
   \colhead{Measured\tablenotemark{a}} &
   \colhead{Ritz\tablenotemark{a}} &
   \colhead{ASD} &
   \colhead{Edl\'en}
   }
\startdata
\ion{Si}{vii} & $2s^22p^4$--$2s2p^5$ &
  $^3P_2$--$^3P_{1}$ & $-2.7105 (9)$ & 272.658 (4) 
  & 272.658 (3 & 272.639 & 272.647\\
&& $^3P_1$--$^3P_{0}$ & $-1.1804 (13)$ & 274.188 (4) 
& 274.188 (4 & 274.175 & 274.180 \\
&& $^3P_2$--$^3P_{2}$ & 0.0000 & 275.368 (4) 
& 275.368 (3 & 275.353 & 275.361\\
&& $^3P_1$--$^3P_{1}$ & $+0.3158 (8)$ & 275.684 (4) 
& 275.685 (3 & 275.667 & 275.675\\
&& $^3P_0$--$^3P_{1}$ & \nodata & \nodata 
& 276.860 (3 & 276.839 & 276.850 \\
&& $^3P_1$--$^3P_{2}$ & $+3.0877 (27)$ & 278.456 (5) 
& 278.456 (3 & 278.443 & 278.449\\
\noalign{\medskip}
\ion{Mg}{vii} & $2s^22p^2$--$2s2p^3$ &
   $^3P_0$--$^3S_{1}$ & $+0.7739 (10)$ & 276.142 (4) 
   & 276.142 (3 & 276.154 & 276.138 \\
&& $^3P_1$--$^3S_{1}$ & \nodata & \nodata  
& 276.993 (3 & 277.001 & 276.993 \\
&& $^3P_2$--$^3S_{1}$ & $+3.0261 (18)$ & 278.394 (4) 
& 278.394 (3 & 278.402 & 278.393\\
&& $^1D_2$--$^1P_1$ & $+5.3613 (14)$ & 280.729 (4) 
& 280.729 (4 & 280.737 & 280.722 \\
\enddata
\tablenotetext{a}{The number in brackets gives the uncertainty on the last digit(s).}
\end{deluxetable}

\section{Derivation of level energies}\label{sect:lopt}

The wavelength measurements presented here allow the energies of the five upper levels of the \ion{Mg}{vii} and \ion{Si}{vii} transitions to be determined when combined with independent measurements of the ground configuration level energies. The wavelengths are input to the publicly available LOPT code \citep{2011CoPhC.182..419K} that was developed at NIST in order to derive an optimized set of level energies from a set of wavelength measurements. Where there are multiple measured wavelengths that constrain a level's energy, the code finds the optimum energy value given the measurement uncertainties. For the present work we consider only the measured wavelengths from the EIS data (given in the fifth column of Table~\ref{tbl.sep}) together with the forbidden transition data discussed below. A comprehensive study of all energy levels that utilizes additional laboratory and space measurements will be performed in a follow-up work. 

Table~\ref{tbl.forbid} lists the measurements of the forbidden transitions within the ground configurations of \ion{Mg}{vii} and \ion{Si}{vii} that are used here. They are obtained from infrared (IR) and far-UV spectra of the Sun and astrophysical nebula sources. \citet{1995ApJ...454L.161K} measured the  \ion{Mg}{vii} $^3P_0$--$^3P_1$ transition, but  the identification was disputed by \citet{1997ApJ...487..962F}, who suggested the line was instead due to \ion{Ar}{iii}. \citet{1999A&A...347..942B} provided a line list for NGC 6302 and confirmed this suggestion. \citet{1999A&A...347..975P} derived a temperature and density for NGC 6302 using emission line diagnostics and the \citet{1999A&A...347..942B} line list. From the CHIANTI 10.1 \ion{Mg}{vii} model \citep{2023arXiv230515221D},  the flux ratio of the 9.0~\um\ line to the 5.5~\um\ line should be 1.16, which then implies that \ion{Mg}{vii} is responsible for 94\%\ of the 9.0~\um\ flux reported by \citet{1999A&A...347..942B}. However, \citet{1999A&A...347..975P} assumed the line was dominated by \ion{Ar}{iii}, which is supported by additional \ion{Ar}{iii} lines, as well as the large fluxes of \ion{Ar}{ii} and \ion{Ar}{v} lines. Inspection of the publicly available \emph{Infrared Space Observatory} spectra of NGC 6302 does not show an obvious alternative identification for the \ion{Mg}{vii} line, but future observations of NGC 6302 with the Mid Infrared Instrument on board the James Webb Space Telescope  may resolve this issue. 

Assuming the \ion{Mg}{vii} 9.0~\um\ line is blended with \ion{Ar}{iii},  an energy of $1112.2\pm 0.2$~cm$^{-1}$ can be assigned to the $^3P_0$--$^3P_1$ transition, corresponding to a wavelength of 8.991~\um. The uncertainty comes from assuming the \ion{Mg}{vii} line is found somewhere within the \ion{Ar}{iii} profile, which is about 0.4~cm$^{-1}$ wide at the base \citep[Figure~2 of][]{1995ApJ...454L.161K}.

The separations of the EUV lines are found to be consistent with the infrared wavelengths. For example, the \ion{Si}{vii} 272.658~\AA\ and 275.685~\AA\ line separation is measured to be $3.0263\pm 0.0011$~\AA. This implies a separation of $4026.1 \pm 1.4$~cm$^{-1}$ for the ground $^3P_{2,1}$ levels, consistent with the infrared measurement (Table~\ref{tbl.forbid}).  The \ion{Mg}{vii} 276.142~\AA\ and 278.394~\AA\ lines imply a separation of $2929.6\pm 1.7$~cm$^{-1}$ for the ground $^3P_{0,2}$ levels. Combining the two IR lines gives a separation of $2929.30\pm 0.46$~cm$^{-1}$, consistent with the EIS measurements.

\begin{deluxetable}{cccccl}[t]
\tablecaption{Forbidden line wavelength measurements.\label{tbl.forbid}}
\tablehead{
\colhead{Ion} &
\colhead{Configuration} &
   \colhead{Transition} &
   \colhead{Wavelength} &
   \colhead{Units} &
   \colhead{Reference}
   }
\startdata
\ion{Si}{vii} & $2s^22p^4$ &
$^3P_2$--$^3P_1$ & $4026.84 \pm 0.20$ & cm$^{-1}$ & \citet{1993A+A...274..662R} \\
&& $^3P_1$--$^3P_0$ & $1540.29 \pm 0.25$ & cm$^{-1}$ & \citet{1997ApJ...487..962F} \\
\noalign{\medskip}
\ion{Mg}{vii} & $2s^22p^2$ &
$^3P_0$--$^3P_1$ & $1112.2 \pm 0.2$ & cm$^{-1}$ & See main text.\\
&& $^3P_1$--$^3P_2$ & $1817.10 \pm 0.41$ & cm$^{-1}$ & \citet{1997ApJ...487..962F}\\
&& $^3P_2$--$^1D_2$ & $2629.674 \pm 0.221$ & \AA\ & \citet{2006ApJ...650.1091Y} \\
\enddata
\end{deluxetable}

Table~\ref{tbl.ens} gives the final energies and uncertainties obtained from LOPT for \ion{Mg}{vii} and \ion{Si}{vii}. They are compared with the values from the NIST ASD, which derive from the compilations of \citet{1980JPCRD...9....1M} and \citet{1983JPCRD..12..323M} for magnesium and silicon, respectively. B.~Edl\'en performed a number of analyzes of energy levels along isoelectronic sequences, and his values for \ion{Mg}{vii} and \ion{Si}{vii} are given in the fourth column of Table~\ref{tbl.ens}. These values were published in \citet{1983PhyS...28...51E} and \citet{1985PhyS...31..345E} for \ion{Si}{vii} and \ion{Mg}{vii}, respectively. The ASD values show differences of up to 26~cm$^{-1}$ compared to the present energies for the levels in the excited configurations. In all cases the Edl\'en energies are closer to those found here. 

The Ritz wavelengths (derived from the optimized level energies) are given in the sixth column of Table~\ref{tbl.sep} and these are the recommended wavelengths to be used when deriving Doppler shifts from EIS. They can be compared with the ASD and Edl\'en wavelengths given in the seventh and eight columns of Table~\ref{tbl.sep}. The ASD wavelengths for \ion{Si}{vii} are systematically smaller by 13 to 21~m\AA, while they are systematically larger for \ion{Mg}{vii} by 8 to 12~m\AA. The Edl\'en wavelengths for \ion{Si}{vii} are also systematically smaller than the present values by 7 to 11~m\AA, but the \ion{Mg}{vii} wavelengths are in good agreement, except for the 280.73~\AA\ line, which is lower by 7~m\AA\ in the Edl\'en work.

\begin{deluxetable}{rccc}
\tablecaption{Comparison of \ion{Si}{vii} and \ion{Mg}{vii} energies with previous values.\label{tbl.ens}}
\tablecolumns{4}
\tablehead{
   & \multicolumn{3}{c}{Energy / cm$^{-1}$} \\
   \cline{2-4}
   \colhead{Level} &
   \colhead{Present\tablenotemark{a}} &
   \colhead{ASD} &
   \colhead{Edl\'en}
   }
\startdata
\sidehead{\ion{Si}{vii} (ground: $^3P_2$)}
$2s^22p^4$ $^3P_1$ & 4026.8 (2) & 4030  & 4028 \\
$^3P_0$ & 5567.1 (2) & 5565  & 5568 \\
$2s^22p^5$ $^3P_2$ & 363150 (4) & 363170 & 363160 \\
$^3P_1$ & 366760 (4) & 366786 & 366774 \\
$^3P_0$ & 368740 (5) & 368761 & 368752 \\
\sidehead{\ion{Mg}{vii} (ground: $^3P_0$)}
$2s^22p^2$  $^3P_1$ & 1112.2 (2) & 1107 & 1118\\
$^3P_2$ & 2929.3 (5) & 2929 & 2933\\
$^1D_2$ & 40956.8 (32) & 40948 & 40957\\
$2s2p^3$ $^3S_1$ & 362132 (4) & 362117 & 362138\\
$^1P_1$ & 397172 (6) & 397153 & 397181  \\
\enddata
\tablenotetext{a}{The uncertainty on the last digit(s) of the energy is given in brackets.}
\end{deluxetable}

\section{Summary and discussion}\label{sect:summary}

The sixth column of Table~\ref{tbl.sep} gives new reference wavelengths for the \ion{Mg}{vii} and \ion{Si}{vii} emission lines observed by \hinode/EIS between 272 and 281~\AA. These values are recommended to EIS users for deriving Doppler shifts velocities for these two ions. The new wavelengths are combined with literature values for the ions' forbidden lines to yield new experimental energies for the transitions' levels (second column of Table~\ref{tbl.ens}).

The new wavelengths should be valuable for the upcoming NASA Multi-slit Solar Explorer \citep[MUSE:][]{2020ApJ...888....3D} mission, which will perform spectroscopy in three EUV channels. One of these is centered at 284~\AA\ and will include the \ion{Mg}{vii} and \ion{Si}{vii} lines discussed in this article. As the instrument uses multiple parallel slits, the resulting spectra will overlap on the detector requiring disambiguation using synthetic spectra \citep{2019ApJ...882...13C}. For the disambiguation to be successful, it is essential that accurate wavelengths are available for the emission lines.

\begin{acknowledgements}
    The author acknowledges funding from the NASA Heliophysics Data Resource Library, the GSFC Internal  Scientist Funding Model competitive work package program, and the \hinode\ project. \hinode\ is a Japanese mission developed and launched by ISAS/JAXA, with NAOJ as domestic partner and NASA and STFC (UK) as international partners. It is operated by these agencies in co-operation with ESA and NSC (Norway).
\end{acknowledgements}

\facility{Hinode}

\bibliography{ms}{}
\bibliographystyle{aasjournal}

\end{document}